\documentclass[twocolumn,trackchanges]{aastex701}
\usepackage{courier}
\usepackage{multirow}
\usepackage{threeparttable}
\usepackage{booktabs}
\usepackage{color}
\usepackage[section]{placeins}
\usepackage{graphicx}
\usepackage{subfigure}
\usepackage{parskip}
\usepackage{longtable}
\usepackage{supertabular}
\usepackage{bm}
\usepackage{amsmath}
\usepackage{amssymb}

\begin{document}

\title{Constraints on the $^{12}$C$(\alpha, \gamma)^{16}$O and $^{16}$O+$^{16}$O Reaction Rates from Binary Black Holes 
Detected via Gravitational Wave Signals}

\author[orcid=0000-0003-3646-9356, sname='Xin']{Wenyu Xin}
\affiliation{School of Physics and Astronomy, Beijing Normal University, Beijing 100875, People's Republic of China}
\affiliation{Institute for Frontiers in Astronomy and Astrophysics,
Beijing Normal University, Beijing 102206, People's Republic of China}
\email[show]{xinwenyu16@mails.ucas.ac.cn}  

\author[0000-0001-9430-2306]{Xiaokun Hou}
\affiliation{Institute of Astronomy and Physics, Inner Mongolia University, Hohhot 010021, People's Republic of China}
\email[]{houxiaokun20@mails.ucas.ac.cn}

\author[0000-0002-3672-2166]{Xianfei Zhang}
\affiliation{School of Physics and Astronomy, Beijing Normal University, Beijing 100875, People's Republic of China}
\affiliation{Institute for Frontiers in Astronomy and Astrophysics,
Beijing Normal University, Beijing 102206, People's Republic of China}
\email[show]{zxf@bnu.edu.cn}

\author[0000-0002-7642-7583]{Shaolan Bi}
\affiliation{School of Physics and Astronomy, Beijing Normal University, Beijing 100875, People's Republic of China}
\affiliation{Institute for Frontiers in Astronomy and Astrophysics,
Beijing Normal University, Beijing 102206, People's Republic of China}
\email[]{bisl@bnu.edu.cn}

\author[0000-0002-8980-945X]{Gang Zhao}
\affiliation{CAS Key Laboratory of Optical Astronomy, National Astronomical Observatories, Chinese Academy of Sciences, Beijing 100101, People's Republic of China}
\affiliation{School of Astronomy and Space Science, University of Chinese Academy of Sciences, Beijing 100049, People's Republic of China}
\email[]{gzhao@nao.cas.cn}


\begin{abstract}
Gravitational-wave observations of binary black hole (BH) mergers
provide a novel avenue for testing massive-star evolution
and the resulting BH mass spectrum.
Recent population analyses under the hierarchical-merger
hypothesis have offered evidence for the BH mass gap and
inferred its lower edge to $\sim 44 - 68$ M$_\odot$.
Motivated by these findings,
we compute low-metallicity ($Z=10^{-5}$) helium star models
with MESA and systematically explore the effect of uncertainties in the 
$^{12}$C$(\alpha, \gamma)^{16}$O and $^{16}$O+$^{16}$O reaction rates
on the final fate. Varying the $^{12}$C$(\alpha, \gamma)^{16}$O
reaction rate by $-3 \sigma$ to $+3\sigma$,
we find that the predicted BH mass gap shifts from $\sim104 - 184$ M$_\odot$
to $\sim45 - 135$ M$_\odot$. 
In contrast, scaling the $^{16}$O+$^{16}$O reaction rate by
global factors of 0.1, 1, and 10 has only a modest effect on the lower edge of the BH
mass gap (less than 5 M$_\odot$), and shifts the upper edge by more than 10 M$_\odot$.
Using the predictions of our models together with the
literature estimates for the lower edge of the BH mass gap,
we constrain the astrophysical S factor of
$^{12}$C$(\alpha, \gamma)^{16}$O
reaction at 300 keV of $S_{300} \simeq$ 137.6 - 263.4 keV barn.
\end{abstract}



\section{Introduction}           
\label{sect:intro}

Gravitational wave (GW) detections have opened a novel avenue for probing
the mass distribution of stellar-mass black holes (BHs)
\citep{2016PhRvL.116f1102A, 2019ApJ...882L..24A}.
Yet, several recent binary BH (BBH) mergers  pose significant
challenges to conventional stellar evolution theory.
The LIGO/Virgo/KAGRA events GW190521 \citep{Abbott_2020} and GW231123
\citep{2025ApJ...993L..25A} originate from mergers involving BHs with component
masses of $85_{-14}^{+21}$ M$_{\odot}$ and $66_{-18}^{+17}$ M$_{\odot}$,
and 137$^{+22}_{-17}$ M$_\odot$ and 103$^{+20}_{-52}$, respectively.
These components are placed at or beyond the conventional pair-instability supernova
(PISN) BH mass gap.

The maximum remnant mass constrains the lower edge of the PISNe BH mass gap 
($\simeq$ 50 M$_\odot$; \citealt{2007Natur.450..390W, 2016A&A...594A..97B})
left by pulsational PISNe (PPISNe; \citealt{1967PhRvL..18..379B}),
which occur for zero-age main-sequence (ZAMS) masses in the range of $\sim$80 - 140 M$_\odot$.
In these stars, explosive O-burning induces multiple pulsations that eject mass
prior to core collapse. The upper edge of the mass 
gap depends on the minimum BH mass ($\simeq 140$ M$_\odot$;
\citealt{2002ApJ...567..532H, 2006ApJ...645.1352O, 2022ApJ...941..100S})
formed by stars with $M(\rm ZAMS) >$ 300 M$_\odot$. 
Between 140 M$_\odot$ and 300 M$_\odot$,
stars are expected to explode as PISNe and leave no remnant
\citep{1967PhRvL..18..379B, 2001ApJ...550..890B, 2002ApJ...565..385U}.
 
The existence of BBHs within or near the mass gap has
spurred extensive work on
how stellar-evolution uncertainties affect its location,
including variations in reaction rates
\citep{2018ApJ...863..153T, Farmer_2019,
Farmer_2020, 2021MNRAS.501.4514C, 2021ApJ...912L..31W, 2022ApJ...924...39M, 2022ApJ...937..112F}; 
rotation \citep{2020A&A...640L..18M, 2021ApJ...912L..31W, 2024MNRAS.529.2980W};
Super-Eddington accretion \citep{2020ApJ...897..100V};
magnetic fields \citep{10.1093/mnras/staa237};
convection and overshooting \citep{Farmer_2019, 2020MNRAS.496.1967K, 2024MNRAS.529.2980W};
stellar merging \citep{2020ApJ...904L..13R}, and metallicity \citep{Farmer_2019, 2024MNRAS.529.2980W}
In particular, numerous studies \citep{Farmer_2019, 
2021MNRAS.501.4514C, 2021ApJ...912L..31W, 2022ApJ...937..112F}
have shown that the uncertainty in the $^{12}$C$(\alpha, \gamma)^{16}$O
reaction rate exerts the most significant impact on the mass gap location. 
A reduction in this reaction rate shifts the gap to higher masses.
Specifically, varying this rate from +3$\sigma$ to $-3 \sigma$,
shifts the mass gap from 43 - 118 M$_\odot$ to 95 - 173 M$_\odot$ \citep{Farmer_2020}. 
Nevertheless, the secondary BH in GW231123 (103 M$_\odot$),
remain within the 95 - 118 M$_\odot$ interval,
which is excluded for all considered reaction-rate values.
Consequently, reconciling this component likely requires
additional degrees of freedom beyond the reaction-rate uncertainties,
such as convective overshooting \citep{2025arXiv250801135T} or rotation \citep{2026MNRAS.546ag073C}.

Beyond the stellar evolution channels, alternative formation scenarios for these
BHs are actively being explored. These include repeated BH mergers
in star clusters \citep{PhysRevD.100.043027, 2021MNRAS.501.5257R, 2025arXiv250717551L},
formation in active galactic nucleus disks \citep{2020ApJ...898...25T},
and primordial BHs from the early matter-dominated epoch \citep{2025arXiv250715701Y}.
Specifically, under the hierarchical merger assumption,
\citet{2025PhRvL.134a1401A} studied the effective inspiral spin ($\chi_{\rm eff}$)
distribution of BHs in the third gravitational wave transient catalog (GWTC-3)
and reported evidence for a BH mass gap, inferring a lower edge of
44$^{+6}_{-4}$ M$_\odot$.
They found that the $\chi_{\rm eff}$ distribution of BHs below this edge, tending to have low spins, can be described by a Gaussian distribution,
while the $\chi_{\rm eff}$ distribution of those above this edge,
characterized by high spins, is better represented by a uniform distribution.
Subsequent analyses based on GWTC-4 have refined the lower edge of the BH mass gap,
yielding values approximately 45, 45.3, 57, and 68.5 M$_\odot$, respectively 
\citep{2025arXiv250904151T, 2025arXiv250904637A, 2025arXiv251018867R,
2025arXiv251022698W}.
These studies also provide constraints on the S‑factor at 300 keV ($S_{300}$)
for the $^{12}$C$(\alpha, \gamma)^{16}$O reaction; however, the reported values
span a wide range from $\sim$101 to 256 keV barns, 
which remain significantly less precise than recent experimental measurements 
\citep{2017RvMP...89c5007D, 2020PhRvL.124p2701S}.

\begin{figure}
\centering
\begin{minipage}[c]{0.5\textwidth}
\includegraphics [width=80mm]{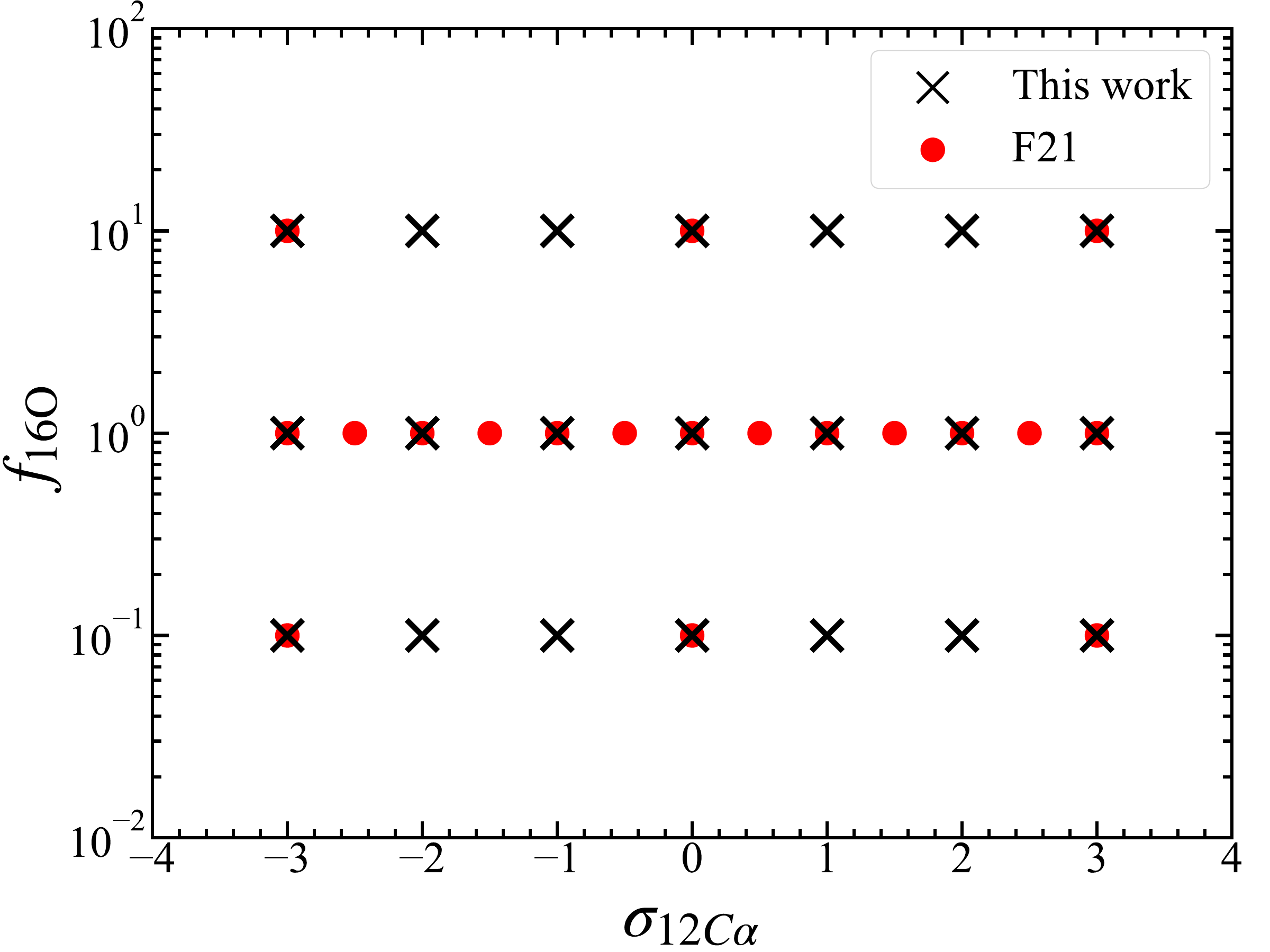}
\end{minipage}%
\caption{The $\sigma_{12C\alpha}$ and $f_{\rm 16O}$ computed in the stellar
models. The red points show the parameter grid computed in \citet{Farmer_2020}
and the black crosses show those computed in this work.
\label{fig:grid}}
\end{figure}

In this work, we investigate the effects of the $^{12}$C$(\alpha, \gamma)^{16}$O
and the $^{16}$O+$^{16}$O reaction rates on the PPISN/PISN with finer grids sampling,
as shown in Figure \ref{fig:grid}.
The inclusion of the $^{16}$O+$^{16}$O reaction is essential because it is expected to
affect the onset of oxygen burning and, consequently, the PPISN/PISN explosion.
We then attempt to constrain the uncertainties of both reaction rates by leveraging
the lower edge of the BH mass gap reported in the literature. The rest of the paper is organized as follows.
In Section \ref{sec:input}, we describe our stellar models and underlying assumptions.
In Section \ref{sec:ppi_bh}, we present the evolution of our models and compare them with
previous studies. In Section \ref{sec:cons}, we apply a Bayesian inference framework
to constrain the two reaction rates based on the outcome of our models.
Finally, in Section \ref{sec:diss} and \ref{sec:con}, we discuss limitations of the present work and summarize our conclusions.

\section{Models and Input Physics} \label{sec:input}

Following \citet{2023RAA....23a5014X}, we employ MESA version r12778
\citet{2011ApJS..192....3P, 2013ApJS..208....4P, 2015ApJS..220...15P, 2018ApJS..234...34P}
to calculate massive He cores with initial masses of $M(\rm He)$ = 30 - 200 M$_\odot$ and
an initial metallicity of $Z$ = 10$^{-5}$.
These models are evolved from the zero-age horizontal branch (ZAHB) through 
the following stages: (1) Fe core collapse (CC) without PPI,
(2) the PPI and then Fe CC (PPISN) and (3) PISNe.

We adopt the same MESA inlists\footnote{We uploaded the related inlist and date files at doi:10.5281/zenodo.18999597} 
and run\_star\_extras.f90 as used in
\citet{Farmer_2020, 2023RAA....23a5014X}. In this work,
we set \texttt{max\_dq} = 10$^{-3}$ and \texttt{mesh\_delta\_coeff} = 0.8 to constrain
the maximum fractional mass of a cell during hydrostatic evolution and
\texttt{split\_merge\_amr\_nz\_baseline} = 6000
which yields 3000 - 4000 cells during hydrodynamic burning.
To ensure a smooth PPISN evolution,
we switch from zoning in log$r$ to zoning in log$\tau$ by disabling
\texttt{split\_merge\_amr\_log\_zoning} and enabling
\texttt{split\_merge\_amr\_logtau\_zoning} for specific models.
The minimum diffusion coefficient of $D_{\rm min}=$ 10$^{-2}$ cm$^2$ s$^{-1}$ 
is adopted to ensure that the global mixing timescale, $\tau=L^2/D_{\rm min}$, 
remains significantly longer than the lifetimes of the stellar models.
This allows us to neglect the effects of global mixing and to smooth local
composition gradients \citep{2022ApJ...937..112F}.
Given the very low metallicity, mass loss driven by winds is negligible \citep{2019ApJ...887...72L, Farmer_2019}.

Previously, \citet{Farmer_2019} found that the size of the
nuclear network had little effect on the final BH masses, and hence,
we employ the \texttt{approx21\_plus\_co56.net}. This network includes the
$\alpha$-chain ($^4$He, $^12$C $^{16}$O, $^{20}$Ne, $^{24}$Mg,
$^{28}$Si, $^{32}$S, $^{36}$Ar, $^{40}$Ca, $^{44}$Ti, $^{48}$Cr, $^{52}$Fe,
and $^{56}$Ni), $^1$H, $^3$He and $^{14}$N for simulating all the nuclear
burning phases. $^{56}$Fe and $^{56}$Co can trace the decay of $^{56}$Ni,
and $^{56}$Cr can be used to keep $Y_{\rm e}$ during the nuclear statistical
equilibrium (NSE) process.
Most reaction rates in this work are drawn from the latest JINA reaclib compilation \citep{2010ApJS..189..240C}.
We use the enhanced rate for the 3$\alpha$ reaction from \citet{2020PhRvL.125r2701K}.
For the $^{12}$C+$^{12}$C reaction, we combine the total rate from
\citet{CAUGHLAN1988283} (CF88) with the branching ratios of the
$\alpha$, p, n channels from \citet{1976NuPhA.265..153D}.
For the $^{12}$C$(\alpha, \gamma)^{16}$O reaction, we adopt a high-resolution table containing approximately
2000 data points, originally provided
by \citet{2017RvMP...89c5007D} and refined in \citet{2022ApJ...924...39M}.
In this work, we vary the $^{12}$C$(\alpha, \gamma)^{16}$O reaction rate
from -3$\sigma$ to +3$\sigma$ with an interval of 1$\sigma$.
Since the temperature-dependent uncertainty is absent in the recent literature about $^{16}$O+$^{16}$O reaction,
we apply a global multiplier $f_{\rm 16O}$ with values of 0.1, 1, and 10.

\section{Evolution of PPISN and the BH mass gap} \label{sec:ppi_bh}

Our calculations cover three cases: $f_{\rm 16O}$ = 0.1, 1, and 10.
In each case, the $^{12}$C$(\alpha, \gamma)^{16}$O reaction rate
is varied
from -3$\sigma$ to +3$\sigma$ with an interval of 1$\sigma$,
resulting in more than 360 model configurations.
The effects of $^{16}$O+$^{16}$O reaction rate on the evolution and
nucleosynthesis of core-collapse supernovae have been discussed in 
\citet{2025arXiv251222963X}, we thus only focus on the PPISN mass range here.
In Section \ref{sec:ppi}, we present the maximum BH masses obtained under our default assumptions.
In Section \ref{sec:bh}, we show the BH mass predictions from this work
and compare with the results in the literature.

\subsection{Maximum BHs}
\label{sec:ppi}

\begin{figure}[htb]
\centering
\begin{minipage}[c]{0.6\textwidth}
\includegraphics [width=80mm]{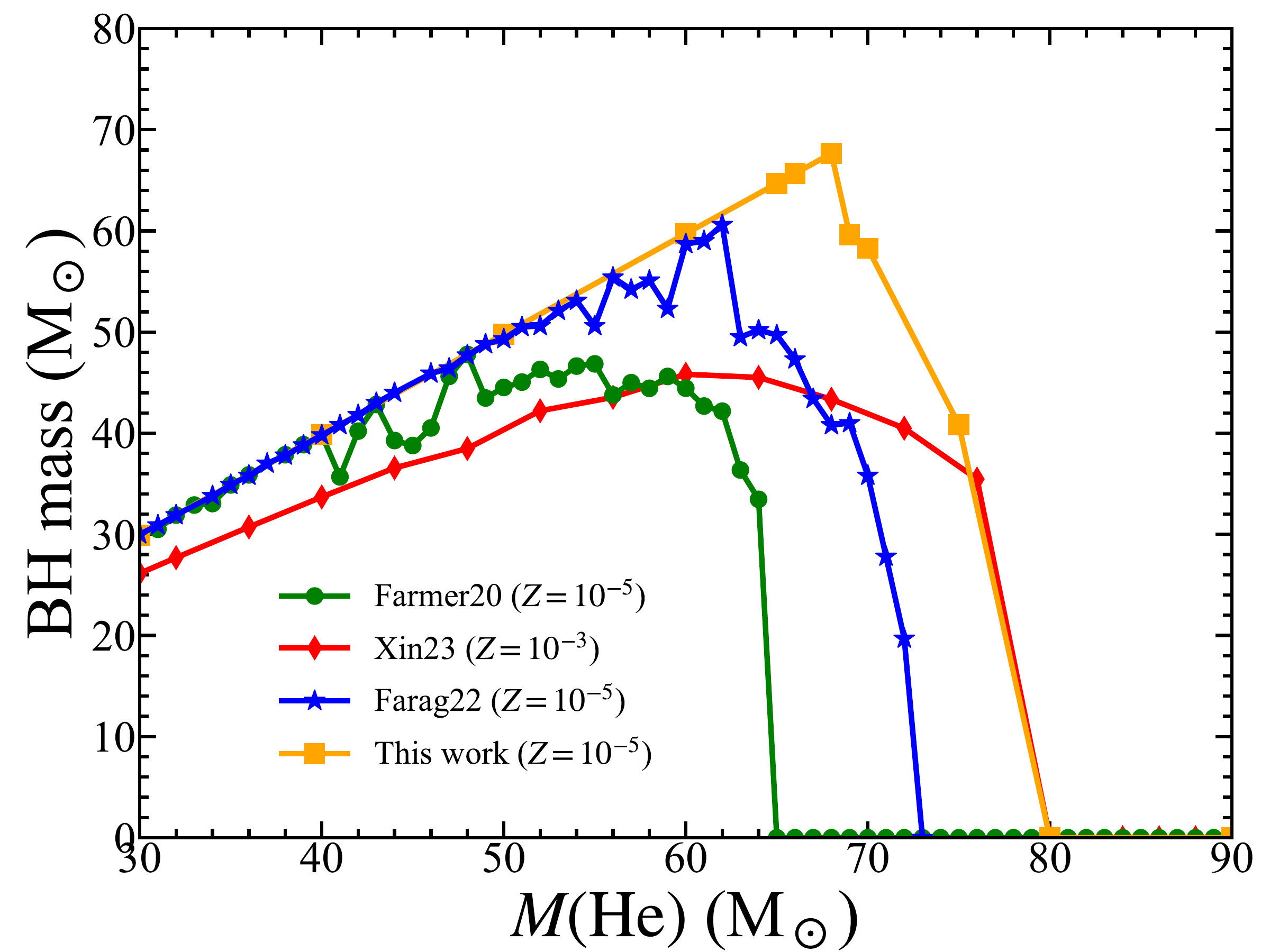}
\end{minipage}%
\caption{The BH masses as a function of initial He core mass.
Green points, blue stars, and red diamonds denote the results from 
\citet{Farmer_2020, 2022ApJ...937..112F} and \cite{2023RAA....23a5014X}.
It should be noted that \cite{2023RAA....23a5014X} adopted $Z=10^{-3}$
while other studies employ $Z=10^{-5}$.
These studies also adopt different reaction rate tables. 
\citet{Farmer_2020} used STARLIB tables for both 3$\alpha$ and
$^{12}$C$(\alpha, \gamma)^{16}$O reaction rates.
\citet{2022ApJ...937..112F} employs NACRE tables for the 3$\alpha$ reaction
but also uses a high-resolution table refined in \citet{2022ApJ...924...39M}.
In \citet{2023RAA....23a5014X}, the 3$\alpha$ reaction rate is from JINA REACLIB,
while the $^{12}$C$(\alpha, \gamma)^{16}$O reaction rate is from
\citet{2002ApJ...567..643K}.
\label{fig:max_bh}}
\end{figure}

Our default He models span initial masses $M(\rm He)$
from 30 M$_\odot$ to 160 M$_\odot$ with $f_{\rm 16O}=1$ and $\sigma_{12C\alpha}=0$.
The properties of these models are listed in Table \ref{tab:ppi},
and the relation between $M(\rm He)$ and $M(\rm BH)$ is compared with
previous studies in Figure \ref{fig:max_bh}.

\begin{table*}[htbp]
\centering
\caption{Behavior of models with our default assumptions. The $M(\rm CO)$ and $X$($^{12}$C)
are defined at the end of He burning. The strong and weak pulses are distinguished by
whether there is mass loss during PPI.}
\label{tab:ppi}
\begin{tabular}{cccccc}
\toprule
$M(\rm He)$ (M$_\odot$)&$M(\rm CO)$ (M$_\odot$)&$X$($^{12}$C)& Pulses &$M(\rm BH)$ (M$_\odot$) & Fate\\
\midrule
30   &    25.95    &    0.2913387097877 &  0                &    29.92    &     CC     \\    
40   &    35.49    &    0.2665980962612 &  0                &    39.86    &     CC     \\    
50   &    44.76    &    0.2089061863652 &  0                &    49.79    &     CC     \\ 
60   &    54.47    &    0.2285757596752 &  0                &    59.72    &     CC     \\
65   &    58.71    &    0.2188018521771 &  0                &    64.68    &     CC     \\
66   &    59.90    &    0.2193547026353 &  0                &    65.67    &     CC     \\
68   &    61.89    &    0.2167031035978 &  1 weak           &    67.65    &     PPISN  \\
69   &    47.81    &    0.2152195001771 &  1 strong, 5 weak   &    59.61  &     PPISN  \\
70   &    63.48    &    0.2129932924089 &  1 strong, 6 weak   &    58.27  &     PPISN  \\
75   &    67.79    &    0.2074153792263 &  1 strong, $\sim$7 weak   &    40.85    &     PPISN  \\
80   &    72.66    &    0.2015297264467 &  1 strong         &    0        &     PISN   \\
90   &    81.42    &    0.1907879364314 &  1 strong         &    0        &     PISN   \\
110  &    99.59    &    0.1734204607957 &  1 strong         &    0        &     PISN   \\
130  &    117.60   &    0.1597809249136 &  1 strong         &    0        &     PISN   \\
135  &    122.27   &    0.1564140820831 &  1 strong         &    0        &     PISN   \\
140  &    126.65   &    0.1539954739087 &  1 strong         &    0        &     PISN   \\
150  &    135.70   &    0.1486403585851 &  1 strong         &    0        &     PISN   \\
151  &    136.65   &    0.1481944843456 &  0                &    149.7    &     CC     \\
152  &    137.45   &    0.1476150373543 &  0                &    150.7    &     CC     \\
153  &    138.52   &    0.1468549235845 &  0                &    151.7    &     CC     \\
155  &    140.22   &    0.1462808071205 &  0                &    153.6    &     CC     \\
160  &    144.74   &    0.1438795416210 &  0                &    158.6    &     CC     \\
\bottomrule
\end{tabular}
\end{table*}

Stars with $M(\rm He) \leq$ 66 M$_\odot$ undergo CC.
The BH mass is defined as the He core mass at CC.
Owing to the very low metallicity, mass loss via stellar winds is
less than 0.5 M$_\odot$ and is negligible.
Although stars with $M(\rm He) =$ 68 M$_\odot$ ignite explosive O burning,
they experience only one weak pulse with no ejecta.
As $M(\rm He)$ increases, the stars undergo one strong pulse followed 
by several weak pulses.
More than 10 M$_\odot$ is expelled during the strong pulse,
causing the BH mass to decrease. Consequently,
the BH mass reaches a maximum at $M(\rm He) =$ 68 M$_\odot$.
This maximum BH mass is larger than those reported in some previous studies.
As shown in Figure \ref{fig:max_bh}, $M(\rm BH_{max})$ is about 46, 47, and 60
M$_\odot$ in \cite{2023RAA....23a5014X}, \citet{Farmer_2020}, and 
\citet{2022ApJ...937..112F}, respectively.
In \cite{2023RAA....23a5014X}, the $M(\rm BH_{max})$ is similar to
that reported in \citet{Farmer_2020}, despite the high progenitor mass.
This difference largely arises because \cite{2023RAA....23a5014X} adopts a metallicity of 
$Z = 10^{-3}$, which reduces the He core mass via stellar winds before explosion. 
The enhancement of $M(\rm BH_{max})$ reported by \citet{2022ApJ...937..112F} results from employing
a high-resolution (2015 data points) table of the $^{12}$C$(\alpha, \gamma)^{16}$O reaction rate.
As noted by \citet{2022ApJ...924...39M}, interpolations based on a low-resolution table
(52 data points) may underestimate this reaction rate during the He burning,
leading to a smaller $X$($^{12}$C) at the end of core He burning.

\begin{figure*}[htbp]
\centering
\begin{minipage}[c]{0.48\textwidth}
\includegraphics [width=85mm]{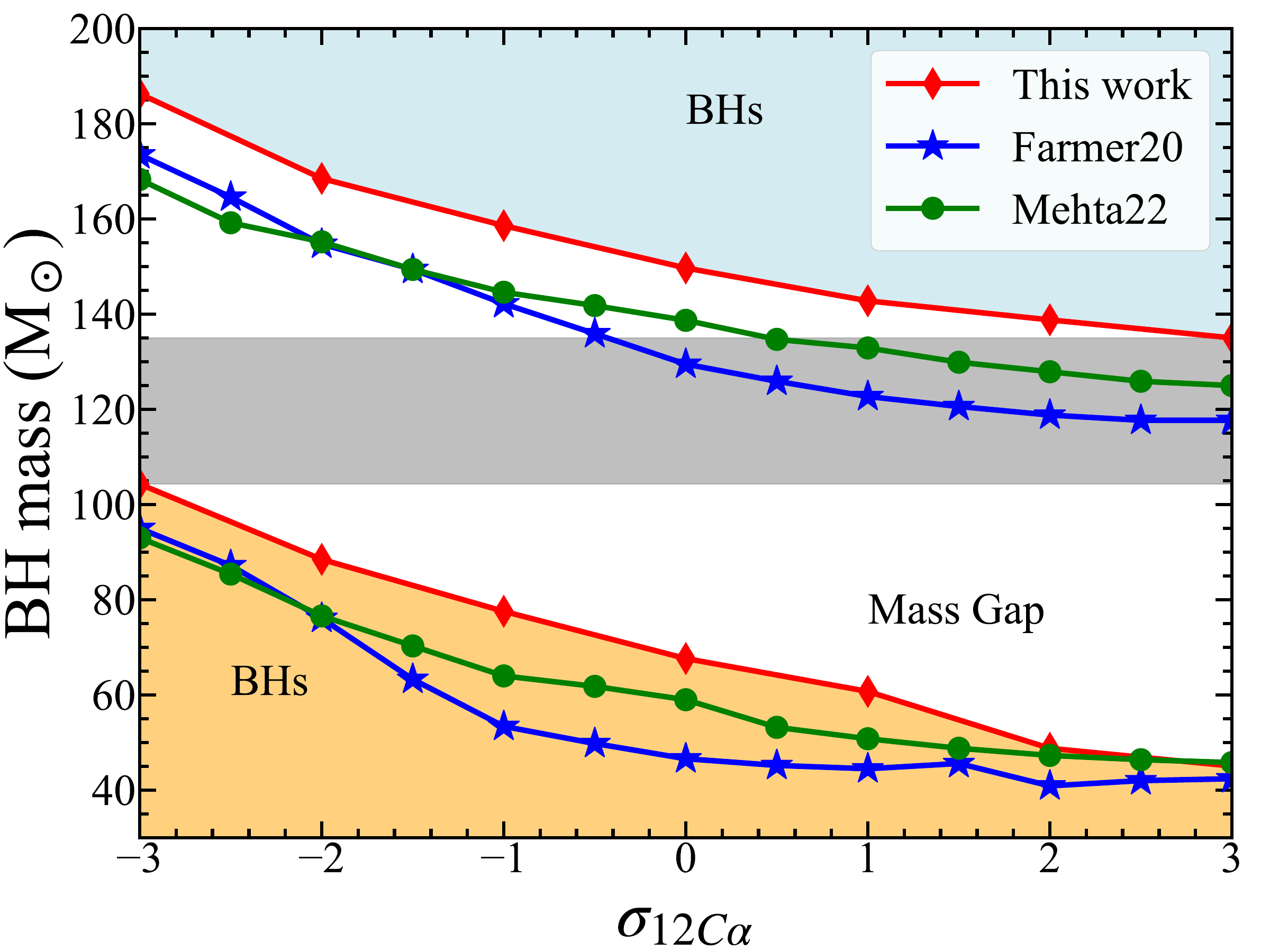}
\end{minipage}%
\begin{minipage}[c]{0.48\textwidth}
\includegraphics [width=85mm]{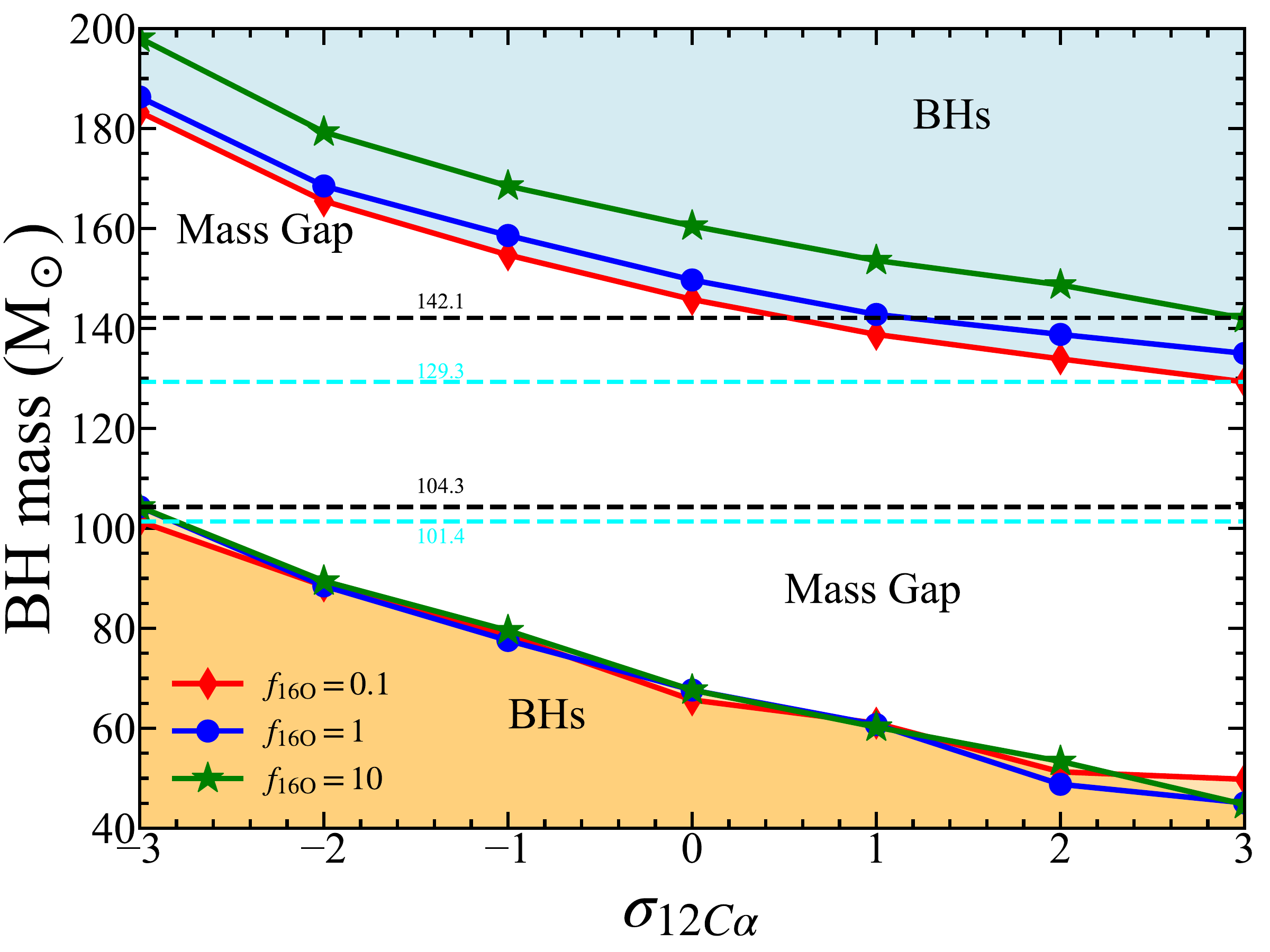}
\end{minipage}%
\caption{The PISN BH mass gap as a function of $\sigma_{12C\alpha}$.
The orange region indicates that the BHs are formed from CCSNe below the
mass gap or PPISNe, while the light blue region represents that the BHs
are formed by the CCSNe above the mass gap. The white region indicates
the mass gap formed due to the PISNe explosion. 
Left: The red, blue, and green lines represent PPISN/PISN and PISN/CC boundaries
from this work, \citet{Farmer_2020}, and \citet{2022ApJ...924...39M}, respectively.
The gray region represents an interval that is excluded
for all considered $\sigma_{12C\alpha}$s.
Right: The red, blue, and green lines represent the results
for $f_{\rm 16O}$ = 0.1, 1, and 10, respectively.
\label{fig:mass_gap}}
\end{figure*}

\begin{table*}[htbp]
\centering
\caption{The mass range investigated in this work and the properties of the mass gap models.}
\label{tab:max_bh}
\begin{tabular}{ccccccc}
\toprule
$f_{\rm 16O}$&$\sigma_{12C\alpha}$&mass range (M$_\odot$)&$M(\rm He)$ (M$_\odot$)&$M(\rm CO)$ (M$_\odot$)&$X$($^{12}$C) &$M(\rm BH_{max})$ (M$_\odot$)\\  
\midrule
 0.1   &  +3  & 30 - 120 & 50  & 41.54 & 0.1134  & 49.79 \\  
 0.1   &  +2  & 40 - 160 & 52  & 46.82 & 0.1601  & 51.28 \\
 0.1   &  +1  & 40 - 140 & 62  & 56.39 & 0.1816  & 61.02 \\  
 0.1   &   0  & 60 - 160 & 66  & 59.74 & 0.2176  & 65.67 \\  
 0.1   &  -1  & 60 - 170 & 79  & 71.64 & 0.2508  & 78.56 \\  
 0.1   &  -2  & 60 - 180 & 89  & 80.25 & 0.2986  & 88.47 \\  
 0.1   &  -3  & 60 - 200 & 102 & 90.90 & 0.3624  & 101.4 \\
\midrule
  1   &  +3  & 30 - 135 & 55  & 49.53 & 0.0036 & 45.07 \\  
  1   &  +2  & 40 - 160 & 49  & 43.40 & 0.1563 & 48.79 \\  
  1   &  +1  & 40 - 160 & 62  & 56.10 & 0.1816 & 60.74 \\     
  1   &   0  & 30 - 160 & 68  & 61.89 & 0.2167 & 67.65 \\          
  1   &  -1  & 40 - 180 & 78  & 70.30 & 0.2517 & 77.57 \\  
  1   &  -2  & 60 - 180 & 89  & 80.45 & 0.2991 & 88.48 \\  
  1   &  -3  & 60 - 200 & 105 & 94.60 & 0.3599 & 104.3 \\ 
\midrule
 10   &  +3  & 30 - 120 & 55  & 49.13 & 0.0034 & 44.75 \\ 
 10   &  +2  & 30 - 180 & 55  & 49.25 & 0.1532 & 53.39 \\  
 10   &  +1  & 40 - 180 & 65  & 59.12 & 0.1768 & 60.23 \\   
 10   &   0  & 60 - 170 & 68  & 61.89 & 0.2167 & 67.65 \\   
 10   &  -1  & 40 - 180 & 80  & 72.20 & 0.2496 & 79.55 \\    
 10   &  -2  & 60 - 200 & 90  & 81.15 & 0.2976 & 89.46 \\  
 10   &  -3  & 60 - 220 & 105 & 94.60 & 0.3599 & 104.3 \\   
\bottomrule
\end{tabular}
\end{table*}

In addition to the high‑resolution table for $^{12}$C$(\alpha, \gamma)^{16}$O
reaction rate, we also adopt an enhanced 3$\alpha$ reaction rate from \citet{2020PhRvL.125r2701K}.
This enhanced rate increases the $X$($^{12}$C) at the end of core He burning.
$X$($^{12}$C) crucially affects the final fate of massive stars
through its effect on shell carbon burning
\citep{2020ApJ...890...43C, 2025arXiv250211012X}.
As \citet{2023RAA....23a5014X} noted,
sufficiently large fuel promotes vigorous shell C burning and
can generate a convective region above the shell.
This convective zone mixes fresh carbon into the base of the shell C burning,
sustaining heating of the outer layers and slowing the contraction of the O-core.
The stars ultimately undergo stable O burning and end as CC.
Conversely, if the fuel is too scant to promote vigorous shell C burning,
the star ignites explosive oxygen burning and undergoes PPISN or PISN.

\subsection{BH Mass Gap and the Effect of $^{16}$O+$^{16}$O Reaction Rate}
\label{sec:bh}

Figure \ref{fig:mass_gap} presents the BH mass gap as a function
of $\sigma_{12C\alpha}$. Under our default assumptions,
the low edge of the BH mass gap is 67.65 M$_\odot$.
Varying $\sigma_{12C\alpha}$ from $-3$ to $+3$ shifts the inferred gap from
104 - 186 M$_\odot$ down to 45 - 135 M$_\odot$.
In the left panel, for $\sigma_{12C\alpha}<2$, our predicted lower edge 
lies above the values reported by \citet{Farmer_2020} and \citet{2022ApJ...924...39M}.
This offset arises from adopting the enhanced 3$\alpha$ reaction rate,
a point we discussed in Section \ref{sec:ppi}.
However, increasing $\sigma_{12C\alpha}$ promotes the conversion of $^{12}$C to $^{16}$O,
which attenuates the impact of the enhanced 3$\alpha$ reaction rate.
In the right panel of Figure \ref{fig:mass_gap},
we compare the BH mass gap for different $f_{\rm 16O}$.
We find that $f_{\rm 16O}$ does not significantly affect the lower edge of
the BH mass gap, with the largest change being less than 5 M$_\odot$. 
By contrast, $f_{\rm 16O}$ noticeably shifts the upper edge of the gap;
in particular, increasing $f_{\rm 16O}$ from 1 to 10 raises the upper edge by
$\sim$11 M$_\odot$.

\section{Constraints the $^{12}$C$(\alpha, \gamma)^{16}$O and $^{16}$O+$^{16}$O reaction rates} \label{sec:cons}
In this section, we present our methodology and results for constraining the
${}^{12}\mathrm{C}(\alpha,\gamma){}^{16}\mathrm{O}$ and
${}^{16}\mathrm{O}+{}^{16}\mathrm{O}$ reaction rates.
Our primary aim is to translate observational constraints on the lower edge of the
BH mass gap into quantitative limits on these nuclear reaction rates.
To achieve this, we first construct a model that replaces direct MESA calculations
and provides a continuous mapping from the reaction rate parameters to the
predicted the lower edge of the mass gap. We then embed this model within a Bayesian
inference framework, wherein the discrepancy between the model-predicted
lower edge and the reference values reported in the literature is used to define the
likelihood and thereby constrain the reaction rates.
This two-step approach provides an efficient and statistically coherent
bridge between stellar evolution calculations and observationally inferred
mass-gap boundaries.

\subsection{RBF model}
\label{sec:4.1}
In Section \ref{sec:ppi_bh}, we perform a suite of MESA models on a discrete 
grid of $^{12}$C$(\alpha, \gamma)^{16}$O and $^{16}$O+$^{16}$O reaction rates.
For each grid point, we determine the lower edge of the mass gap, denoted by $m$.
Because it is not feasible to exhaustively evaluate $m$ over the continuous
reaction rate space, we construct a model mapping from the reaction rates
to $m$ using a two-dimensional radial basis function (RBF):
\begin{equation}
    m = f\!\left(\sigma_{12C\alpha},\,\sigma_{16O}\right)
\end{equation}
here the two inputs $\sigma_{12C\alpha}$ and $\sigma_{16O}$ have different meanings.
The parameter $\sigma_{12C\alpha}$ quantifies the deviation of the
${}^{12}\mathrm{C}(\alpha,\gamma){}^{16}\mathrm{O}$ reaction rate from its
reference value in units of the quoted uncertainty,
whereas for ${}^{16}\mathrm{O}+{}^{16}\mathrm{O}$ we vary the rate by a
multiplicative factor $f_{16O}$. For numerical convenience,
we re-parameterize this factor as $\sigma_{16O}\equiv \log f_{\rm 16O}$.

\begin{figure*}[htb]
    \centering
    \includegraphics[width=0.8\textwidth]{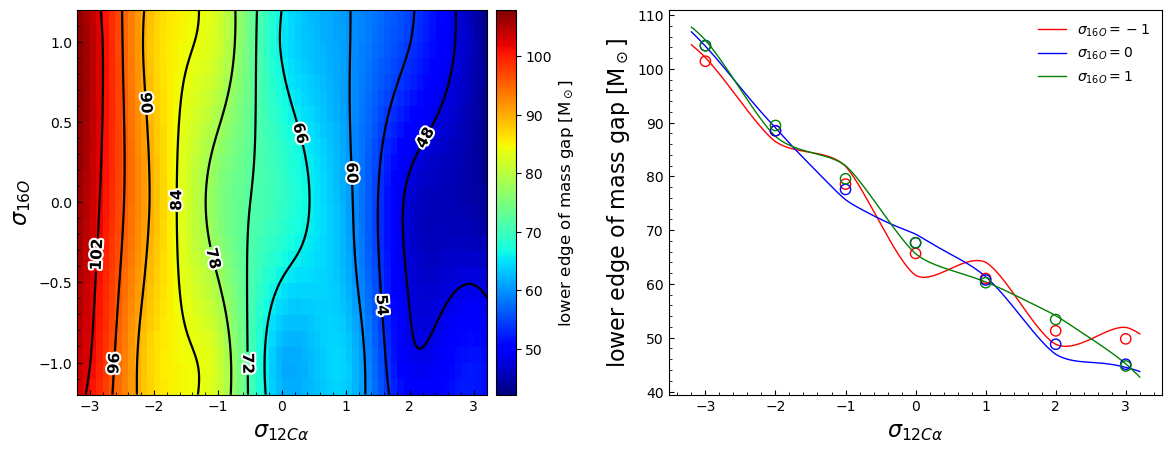}
    \caption{
    Performance of the RBF model as a computational substitute for the discrete MESA calculations in determining the lower edge of the mass gap. 
    \textit{Left panel:} Color map of the RBF-predicted $m$ over the $\sigma_{12C\alpha}-\sigma_{16O}$ space. Black contour lines denote iso-$m$ levels. 
    \textit{Right panel:} One-dimensional slices of the RBF surface at fixed values of $\sigma_{16O}=-1,\,0,$ and $1$. Solid curves represent the RBF predictions, while symbols show the corresponding MESA grid results. 
}
    \label{fig:rbf-contour}
\end{figure*}

The RBF model is implemented with the \texttt{Rbf} routine in \texttt{SciPy}. We adopt a thin-plate spline kernel and allow a non-zero smoothing parameter (\texttt{smooth}$=0.3038$), so the fitted surface is not required to pass exactly through every MESA grid point. This choice is motivated by the sparsity of the MESA grid: enforcing exact interpolation (\texttt{smooth}$=0$) can lead overfit and introduce spurious local oscillations between grid points. Using a non-zero smoothing parameter yields a more stable surrogate surface and improves the robustness of subsequent inference based on the RBF-predicted $m$. Fig. \ref{fig:rbf-contour} illustrates the performance of the RBF model as a computationally efficient substitute for the discrete MESA calculations in determining the lower edge of the mass gap, over a large reaction-rate space. In the left panel, the color map shows the RBF-predicted $m$ across the $\sigma_{12C\alpha}-\sigma_{16O}$ plane, with warmer colors indicating larger values of $m$ and cooler colors corresponding to smaller values. Black contour lines are plotted to highlight iso-$m$ levels, demonstrating a smooth and predominantly monotonic dependence of $m$ on $\sigma_{12C\alpha}$, while the variation along the $\sigma_{16O}$ direction is comparatively weaker. The right panel presents one-dimensional slices of the RBF surface at fixed values of $\sigma_{16O}=-1,\,0,$ and $1$. Solid curves denote the RBF predictions, and the symbols show the corresponding MESA grid points. The close agreement between the curves and the discrete MESA results indicates that the RBF model captures the overall trend of $m$ while providing a smooth continuous mapping suitable for the subsequent Bayesian inference.

\subsection{Constrain reaction rates with Bayesian model}
In this study, we employ a Bayesian inference framework to constrain the $^{12}$C$(\alpha, \gamma)^{16}$O and $^{16}$O+$^{16}$O reaction rates. We define the parameter vector as:
\begin{equation}
    \boldsymbol{\theta} = \begin{bmatrix} \sigma_{12C\alpha} \\ \sigma_{16O} \end{bmatrix},
\end{equation}
and assign a bi-variate normal prior:
\begin{equation}
    P(\boldsymbol{\theta})
    =
    \frac{1}{(2\pi)^{k/2}\,|\boldsymbol{\Sigma}_{\mathrm{prior}}|^{1/2}}
    \exp\!\left[
    -\frac{1}{2}
    \left(
    \boldsymbol{\theta}-\boldsymbol{\mu}_{\mathrm{prior}}
    \right)^{\mathsf T}
    \boldsymbol{\Sigma}_{\mathrm{prior}}^{-1}
    \left(
    \boldsymbol{\theta}-\boldsymbol{\mu}_{\mathrm{prior}}
    \right)
    \right],
\end{equation}
where $k=2$, and the prior mean and covariance are set to:
\begin{equation}
    \mu_{\text{prior}} = \begin{bmatrix} 0.0 \\ 0.0 \end{bmatrix}, \quad \Sigma_{\text{prior}} = \begin{bmatrix} 1.0^2 & 0.0 \\ 0.0 & 1.0^2 \end{bmatrix}
\end{equation}

We then define the likelihood using the discrepancy between the lower edge of the mass gap predicted by the RBF model and the reference value reported in the literature:
\begin{equation}
    L(m_0 \mid \boldsymbol{\theta})
    =
    \frac{\sqrt{2/\pi}}{\sigma_{+}+\sigma_{-}}
    \exp\!\left[
    -\frac{1}{2}
    \left(
    \frac{m(\boldsymbol{\theta})-m_0}{\sigma}
    \right)^2
    \right],
\end{equation}
where $m_0$ denotes the literature value of the lower edge of the mass gap, and $\sigma_{+}$ and $\sigma_{-}$ represent the corresponding upper and lower uncertainties reported in the same work. The quantity $m(\boldsymbol{\theta})$ is the RBF-predicted lower edge of the mass gap at a given $\boldsymbol{\theta}$. The effective uncertainty $\sigma$ is defined in a piecewise manner:
\begin{equation}
    \sigma =
    \begin{cases}
    \sigma_{+}, & m(\boldsymbol{\theta}) \ge m_0, \\
    \sigma_{-}, & m(\boldsymbol{\theta}) < m_0 .
    \end{cases}
\end{equation}

This formulation explicitly incorporates the asymmetric mass uncertainties reported in the literature. Owing to its exponential form, the likelihood is maximized when the RBF-predicted mass approaches the reference value and decreases rapidly as their discrepancy grows.

\begin{figure}[htb]
\centering
\includegraphics[width=0.48\textwidth]{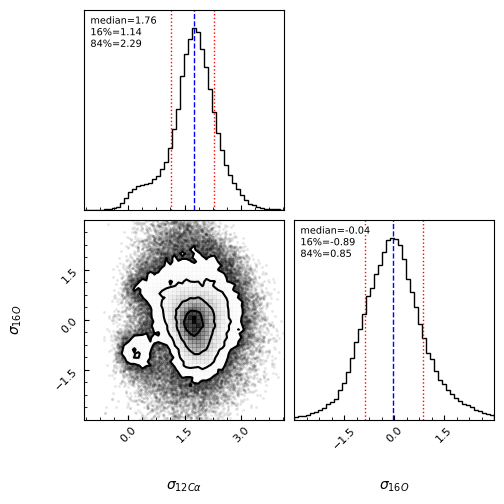}
\caption{Posterior distribution of the $^{12}$C$(\alpha, \gamma)^{16}$O and $^{16}$O+$^{16}$O
reaction rates, inferred using the lower edge of mass gap from \cite{2025arXiv250904637A}
($m_0 = 45.3^{+6.5}_{-6.8}$).
The top and right panels show the marginalized posterior distributions of
$\sigma_{12C\alpha}$ and $\sigma_{16O}$, respectively, while the lower-left panel
displays their joint posterior distribution. The blue dashed lines indicate the
posterior medians, and the red dashed lines mark the 16th and 84th percentiles.}
\label{fig:antonini}
\end{figure}

The posterior function is obtained by multiplying the prior distribution with the likelihood function,
\begin{equation}
P(\boldsymbol{\theta}\mid m_0)
\;\propto\;
P(\boldsymbol{\theta})\,
L(m_0\mid \boldsymbol{\theta}),
\end{equation}

We sampled the posterior function to estimate the reaction rates.
As an illustrative example, we adopt the lower edge of the mass gap reported by
\cite{2025arXiv250904637A} ($m_0 = 45.3^{+6.5}_{-6.8}$), 
the resulting distribution of the samples is shown in Fig. \ref{fig:antonini}. 
From these samples, we find a posterior median of $\sigma_{12C\alpha}$ at 1.76 
with 16th and 84th percentiles at 1.14 and 2.29, and a posterior median of 
$\sigma_{16O}$ at -0.06 with the 16th and 84th percentiles at -0.90 and 0.84.
In addition to \cite{2025arXiv250904637A}, we also consider the lower edge of 
the mass gap from \cite{2025arXiv250904151T}, \cite{2025arXiv251018867R},
and \cite{2025arXiv251022698W}. The corresponding posterior summaries are
listed in Table \ref{tab:suggested}. It is clear that $\sigma_{12C\alpha}$
is strongly correlated with the adopted lower edge of the mass gap,
whereas the median of $\sigma_{16O}$ is largely insensitive to it.
This behavior is consistent with our analysis in Section \ref{tab:suggested},
which indicates a weak dependence of the lower edge of the mass gap on $\sigma_{16O}$.
Consequently, $\sigma_{16O}$ has only a minor impact on the likelihood,
and with a prior centered at 0, its posterior median remains close to zero.

\begin{table*}[htp]
\centering
\caption{$m_{0}$ represents the lower edges of the BH mass gap adopted from the Reference.
The suggested $\sigma_{12C\alpha}$ and $\sigma_{16O}$ are the posterior median together
with the 16th (lower limit) and 84th (upper limit) percentiles. The default $S_{300}$
used in our MESA models is 140 keV barn from \citet{2017RvMP...89c5007D}.}
\label{tab:suggested}
\begin{tabular}{lccccc}
\toprule
Reference&$m_{0}$ (M$_\odot$)&  $\sigma_{12C\alpha}$&  $S_{300}$ (keV barn)&   $\sigma_{16O}$ &  $f_{\rm 16O}$\\
\midrule
\cite{2025arXiv250904151T}&$45.0^{+5.0}_{-4.0}$&1.98$^{+0.48}_{-0.37}$&263.4$^{+48.8}_{-32.2}$&$-0.09^{+0.73}_{-0.71}$  & $0.81^{+3.55}_{-0.65}$ \\
\cite{2025arXiv250904637A}&$45.3^{+6.5}_{-6.8}$&1.76$^{+0.53}_{-0.62}$&243.7$^{+50.2}_{-47.3}$&$-0.04^{+0.89}_{-0.85}$  & $0.87^{+6.05}_{-0.75}$ \\ 
\cite{2025arXiv251018867R}&$57.0^{+17.0}_{-10.0}$&0.26$^{+0.87}_{-0.81}$&146.6$^{+49.2}_{-32.4}$&$-0.03^{+1.01}_{-0.99}$ & $0.93^{+8.62}_{-0.84}$ \\   
\cite{2025arXiv251022698W}&$65.8^{+19.8}_{-18.5}$&0.06$^{+0.91}_{-0.89}$&137.6$^{+47.8}_{-32.4}$&$0.01^{+0.98}_{-0.98}$ & $1.02^{+8.75}_{-0.92}$ \\   
\bottomrule
\end{tabular}
\end{table*}

\section{Discussion} \label{sec:diss}

The results presented in this work rest on the following assumptions.
First, we assume that the BHs detected by LIGO/Virgo/KAGRA originate from stars
that lose their entire hydrogen envelopes before collapsing into BHs.
Consequently, the maximum BH mass is limited by the PISN explosions and
the mass loss associated with PPISNe.
If these stars were to retain their hydrogen envelopes due to the weak stellar wind
\citep{2026arXiv260102263R}, the situation would become more complex 
\citep{2019ApJ...887...72L, 2021MNRAS.501.4514C, 2025A&A...703A.215S}.
Second, the effect of rotation is not taken into consideration in this study
because the first-generation BHs are assumed to be low-spin
\citep{2025PhRvL.134a1401A, 2025arXiv250904151T, 2025arXiv250904637A, 
2025arXiv251018867R, 2025arXiv251022698W}.
However, rotation should play a crucial role in the ultimate fate of massive stars.
\citet{2021ApJ...912L..31W} has reported that the maximum BH mass is enhanced by
$\sim20\%$ to increase the initial total angular momentum of the helium star from 
1.5$\times10^{52}$ erg s to 12 $\times10^{52}$ erg s.
Third, only two reaction rates: $^{12}$C$(\alpha, \gamma)^{16}$O and $^{16}$O+$^{16}$O
are considered in this work. However, there are still many reaction rates that affect
the evolution of massive stars \citep{2016MNRAS.463.4153R, 2018ApJS..234...19F, 
2023ChPhC..47c4107X, 2025arXiv251222963X}, for example, 
$3\alpha$, $^{12}$C+$^{16}$O, and $^{12}$O+$^{12}$O.
Fourth, \citet{Farmer_2019} indicated that the effects of the efficiency
parameter of mixing length theory, $\alpha_{\rm MLT}$, and overshooting on the maximum
$M(\rm BH)$ are limited. The variations induced by these parameters are smaller than 2 M$_\odot$.
By contrast, inefficient convection and overshooting during hydrogen burning, as discussed in 
\citet{2021MNRAS.501.4514C} and \citet{2025arXiv250801135T}, can reduce $M(\rm He)$ and
thus affect $M(\rm BH)$. The discrepancy between these two results largely hinges on
whether the hydrogen envelope is included in $M(\rm BH)$. If one assumes that the H
envelope material also collapses into the BH, then even a small He core combined with
a substantial H envelope could yield a larger $M(BH)$.
Conversely, under the assumption that the H envelope will be lost entirely before collapse, 
as \citet{Farmer_2019, Farmer_2020, 2021ApJ...912L..31W}, and this study,
then the impact of convection and overshooting remains limited. 
Fifth, \citet{Farmer_2019} also indicated that the effects of spatial and temporal
resolution are limited, and these effects on the peak of the BH mass spectra are
further investigated by \citet{2022ApJ...937..112F}.

In future studies, we will extend the grid of models we investigate.
We will also incorporate rotational effects and the 3$\alpha$ reaction rate,
and discuss the constraints on both the 3$\alpha$ and $^{12}$C$(\alpha, \gamma)^{16}$O
reaction rates under different initial orbital angular momenta.

\section{Conclusion} \label{sec:con}

Gravitational-wave observations of binary black hole (BH) mergers
provide a new avenue to test massive-star evolution and the resulting
BH mass spectrum. Recent population analyses under the hierarchical-merger
assumptions have provided a clue to the existence of the BH mass gap and
inferred its lower edge to $\sim 44 - 68$ M$_\odot$.
In this paper, we calculate a series of massive He cores with an initial
mass of $M(\rm He)$ = 30 - 200 M$_\odot$ and an initial metallicity of $Z$ = 10$^{-5}$.
These models are evolved from the zero-age horizontal branch (ZAHB) and end their lives
as CC, PPISN, or PISN.

Based on these models, we systematically assess the impact of uncertainties in the 
$^{12}$C$(\alpha, \gamma)^{16}$O and $^{16}$O+$^{16}$O reaction rates on the final 
fates of massive He stars and on the resulting location of the PISN BH mass gap.
Our principal conclusions are:

(1) The mass-gap lower edge is dominated by $^{12}$C$(\alpha, \gamma)^{16}$O reaction.
Across our models, changing $\sigma_{12C\alpha}$ from $-3$ to $+3$ shifts the predicted BH
mass gap from $\sim104 - 184$ M$_\odot$ down to $\sim45 - 135$ M$_\odot$. 

(2) $^{16}$O+$^{16}$O reaction has a limited impact on the lower edge but affects the upper edge.
Varying a global scaling factor of $f_{\rm 16O}$ from 0.1 to 10, the lower edge of the BH mass gap
only changes by no more than 5 M$_\odot$, while the upper edge is shifted by 13 - 15 M$_\odot$.

Then, we constructed a two-dimensional radial basis function (RBF) surrogate,
$m=f(\sigma_{12C\alpha}, \sigma_{16O})$,
to map reaction-rate parameters to the predicted lower edge of the BH mass gap.
We embedded this surrogate within a Bayesian inference framework,
using the lower edge of the BH mass estimated from binary BH (BBH) detections to
define the likelihood.
However, published estimates of the lower edge differ substantially across the
literature. Consequently, the $S_{300}$ inferred in this work still spans a broad
range from 137.6 to 263.4 keV barn.

\begin{acknowledgments}
This work is supported by the National Natural Science Foundation of China
(Nos. 12588202, 12473028, 12073006, 12090040, and 12090042).
W. Y. X. is supported by the Cultivation Project for LAMOST Scientific Payoff
and Research Achievement.
\end{acknowledgments}


\bibliography{sample701}{}
\bibliographystyle{aasjournalv7}



\end{document}